\documentstyle[12pt,twoside,fleqn,espcrc1,epsf]{article}

\title{Search for a $\Theta^{++}$ Pentaquark State}

\author{H. G. Juengst\address[GWU]{The George Washington University}\\
For the CLAS Collaboration}

\begin{document}


\newcommand{\setnormallineskip}{}%

\setnormallineskip

\renewcommand{\topfraction}{.9999}%
\renewcommand{\bottomfraction}{.0001}%
\renewcommand{\textfraction}{.0001}%
\renewcommand{\floatpagefraction}{.0001}%
\renewcommand{\dbltopfraction}{.9999}%
\renewcommand{\dblfloatpagefraction}{.0001}%
\setcounter{bottomnumber}{99}
\setcounter{topnumber}{99}
\setcounter{dbltopnumber}{99}
\setcounter{totalnumber}{99}

\maketitle

\begin{abstract}
Recent reports about the $\Theta^+$ (formerly known as Z$^+$), a pentaquark state, have awakened considerable interest. We report
our findings in search of a $\Theta^{++}$ state over a wide range in mass in the reaction $\gamma$p$\rightarrow\Theta^{++}$K$^-$
from our analysis of CLAS data. The existence or nonexistence of this state might resolve the question of the isotensor states.
\end{abstract}

\section{The search for the $\Theta^{++}$}

The constituent quark model does not exclude pentaquark systems. Until recently, only bound triplet-quark and quark-antiquark
systems have been found. A possible candidate for a pentaquark system is the $\Theta^{+}$, with strangeness S=+1. Recent
experimental evidence \cite{thet1,thet2,thet3,thet4,thet5,thet6,thet7} may suggest the existence of such a pentaquark system,
previously predicted by the chiral soliton approach to hadron dynamics \cite{Diak97}. Thus the search is on for a family of
pentaquark systems. Theoretical predictions of a family, including $\Theta^{-}$, $\Theta^{0}$, $\Theta^{+}$, $\Theta^{++}$, and
$\Theta^{+++}$, have been published, for instance \cite{Caps03,BinW03,Elli04}.

In 2000, we analyzed the CLAS g1a data \cite{CLAS03} and searched for the $\Theta^{++}$ in an exclusive measurement of the reaction
$\gamma$p$\rightarrow$K$^+$K$^-$p, as a by-product of a hyperon-resonance analysis \cite{Juen01}. The $\Theta^{++}$ may be found in
the reaction $\gamma$p$\rightarrow$K$^-\Theta^{++}$ with an assumed strong decay $\Theta^{++}\rightarrow$K$^+$p. Thus, we require
the exclusive identification of all three particles in the final state, namely, p, K$^+$, and K$^-$. We searched for a structure in
the mass distribution between 1.5 GeV/c$^2$ and 1.6 GeV/c$^2$. The range is based on theoretical predictions. The only structure
that we found in this mass region was at 1.59 GeV/c$^2$ which we label ``$\Theta^{++}$''. We then tried to understand the origin of
this structure. After evidence for a $\Theta^{+}$ was found and the physics community was more interesting in the $\Theta^{++}$, we
received several requests that encouraged us to present the result of this analysis in light of recent findings.

The photon beam energy range of the g1a data set is E$_\gamma$ = 0.5 GeV to 2.4 GeV. The threshold for the reaction
$\gamma$p$\rightarrow$K$^-\Theta^{++}$ is E$_\gamma$ = 1.85 GeV, assuming a mass $m_{\Theta^{++}} = 1.59$ GeV/c$^2$. Here we take
the peak center position of the $\Theta^{++}$ mass distribution which we find below. In the first iteration, no cut was applied on
E$_\gamma$, but for the final analysis only events above the threshold are selected. Even if the $\Theta^{++}$ mass were to differ
by a few tens of MeV, because we cover a wide energy range up to about 0.5 GeV above the threshold, the effect on the mass
distributions shown in this paper is negligible. For the search for the $\Theta^{++}$, the energy range covered by the measurement
presented in this paper is most interesting, because the cross section for the reaction $\gamma$p$\rightarrow$K$^-\Theta^{++}$ would
very likely reach its maximum within this range. We discuss the $\Theta^{++}$ mass below.

\begin{figure}[ht]
  \epsfbox{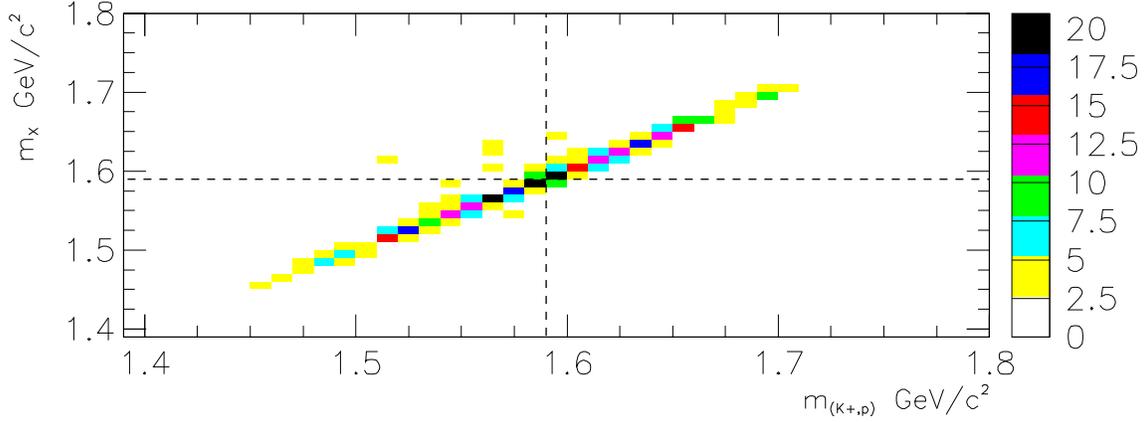}
  \caption{Missing mass $m_X$ vs. inv. mass $m_{(K^+,p)}$ for $\gamma$p$\rightarrow$K$^-$X and
   $\gamma$p$\rightarrow$K$^-$K$^+$p from events with p, K$^+$ and K$^-$ in the final state, whereby particles may be
   misidentified}
  \label{mm_vs_im_1}
\end{figure}

\begin{figure}[ht]
  \epsfbox{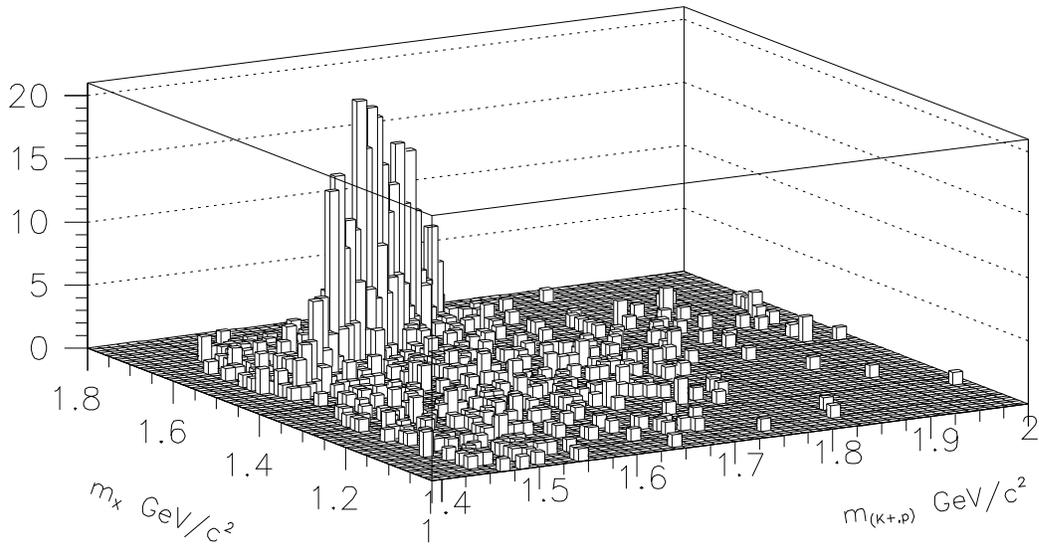}
  \caption{Missing mass $m_X$ vs. inv. mass $m_{(K^+,p)}$ for $\gamma$p$\rightarrow$K$^-$X and
   $\gamma$p$\rightarrow$K$^-$K$^+$p from events with p, K$^+$ and K$^-$ in the final state, same as Fig.
   \protect\ref{mm_vs_im_1}, but showing better the background from particle misidentification below the diagonal}
  \label{mm_vs_im_2}
\end{figure}

\begin{figure}[ht]
  \epsfbox{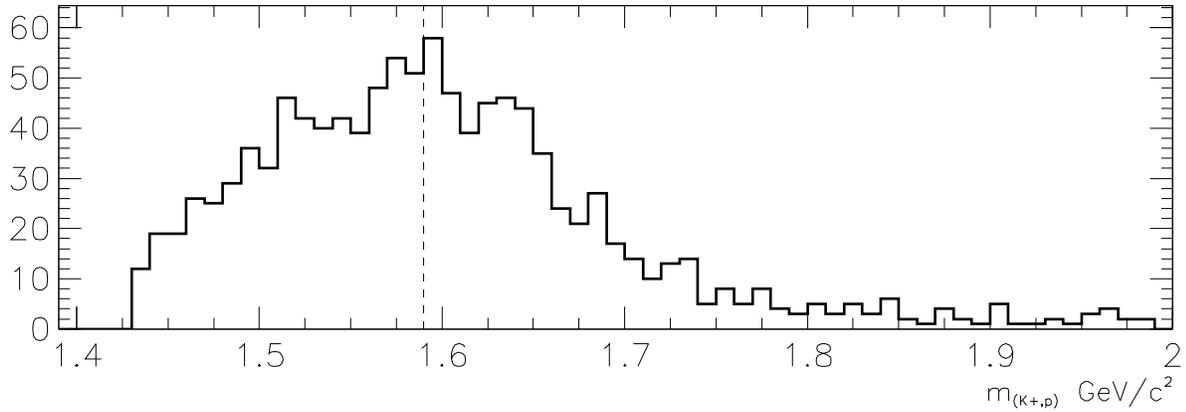}
  \caption{Invariant mass $m_{(K^+,p)}$ for $\gamma$p$\rightarrow$K$^-$K$^+$p from events with p, K$^+$ and K$^-$ in the final
   state, X--projection of Fig. \protect\ref{mm_vs_im_1}}
  \label{mm_vs_im_3}
\end{figure}

\begin{figure}[ht]
  \epsfbox{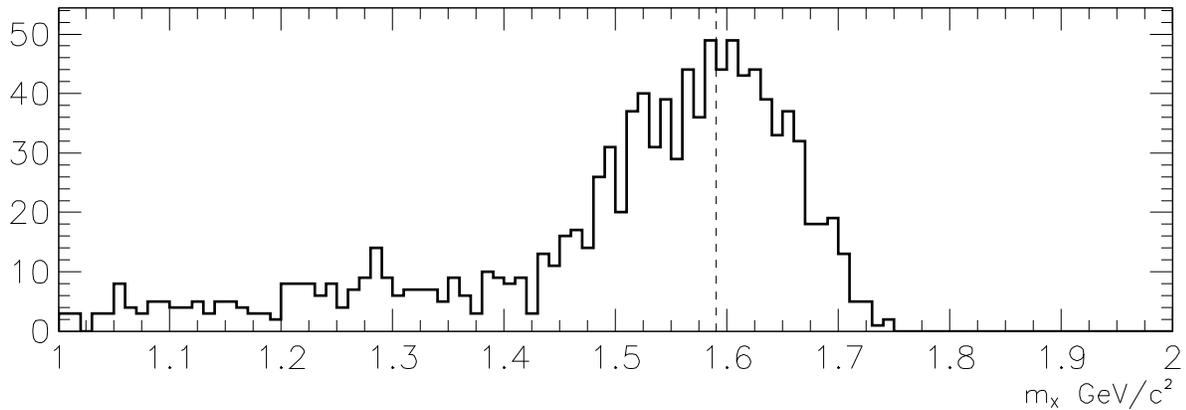}
  \caption{Missing mass $m_X$ for $\gamma$p$\rightarrow$K$^-$X from events with p, K$^+$ and K$^-$ in the final state,
   Y--projection of Fig. \protect\ref{mm_vs_im_1}}
  \label{mm_vs_im_4}
\end{figure}

Figure \ref{mm_vs_im_1} shows the missing mass $m_X$ from $\gamma$p$\rightarrow$K$^-$X versus the invariant mass $m_{(K^+,p)}$ from
$\gamma$p$\rightarrow$K$^-$K$^+$p for the selected events. The same distribution is presented in Fig. \ref{mm_vs_im_2} lego plot
format. The X-- and Y--projection of this distribution are shown in Fig. \ref{mm_vs_im_3} and \ref{mm_vs_im_4}. Notice that the
missing mass $m_X$ in Fig. \ref{mm_vs_im_4} comes from the p and $K^+$ since both particles are required by the event selection
although they are not used for the calculation of $m_X$. A bump can be seen at about 1.59 GeV/c$^2$ in Fig. \ref{mm_vs_im_1},
\ref{mm_vs_im_3} and \ref{mm_vs_im_4}. The missing mass distribution contains a flat background, wherein some events have an
occasional missing mass $m_X$ below the total of m$_{{\rm K}^-}$ + m$_{{\rm K}^+}$ + m$_{\rm p}$. These events most likely arise
from particle misidentification. We can also see these event in Fig. \ref{mm_vs_im_2} below the diagonal. There are two groups of
background reactions, namely $\gamma$p$\rightarrow\pi^-$K$^+$p from $\Lambda$ and $\Sigma^0$ production and
$\gamma$p$\rightarrow\pi^-\pi^+$p. The cross sections for these reactions are significantly larger than the cross section for
double-kaon production. Thus, the result is not surprising. This is the most important reason to make an exclusive measurement and
not to rely on the missing mass distribution only.

We utilize four-momentum conservation as an additional event selection criterion. Because this is an exclusive measurement, it is
possible to look at the missing mass $m_X$ as well as the invariant mass $m_{(K^+,p)}$ without fixing the mass of the $\Theta^{++}$.
We merely require that the missing mass and invariant mass be identical within some reasonable range. This is achieved by a cut on
$|m_X-m_{(K^+,p)}| \le 35$ MeV/c$^2$. In Fig. \ref{mm_vs_im_1} and \ref{mm_vs_im_2}, which show $m_X$ plotted against $m_{(K^+,p)}$,
this would be a 35 MeV/c$^2$ broad diagonal cut. This cut removes the background which does not conserve the four-momentum and is
not on the diagonal. No cut on the particle momentum or any other kinematic cut has been applied.

\begin{figure}[ht]
  \epsfbox{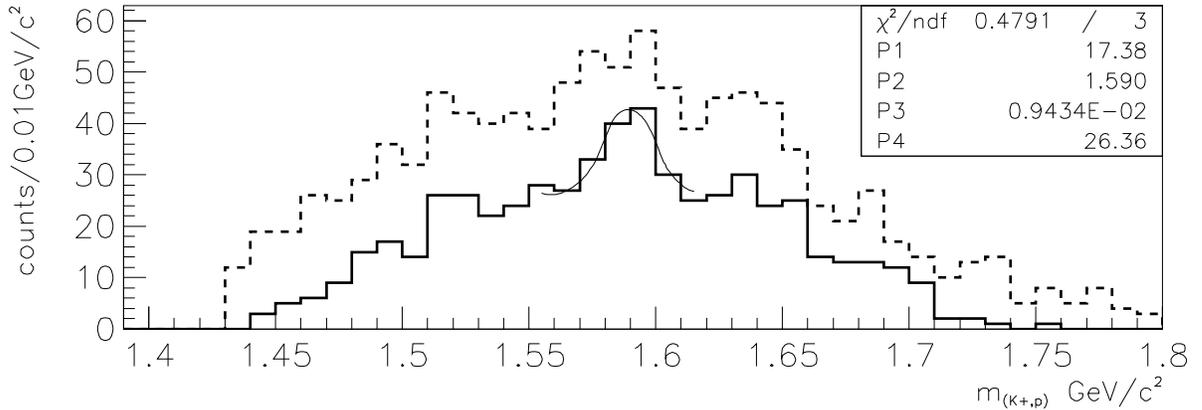}
  \caption{Improved signal of the $\Theta^{++}$ candidates; invariant mass $m_{(K^+,p)}$ with (solid line) and without
   (dashed line) the cut on $|m_X-m_{(K^+,p)}| \le 35$ MeV/c$^2$, fit of a normal distribution plus a constant offset to the
   $\Theta^{++}$ signal, the numbers in the box are the fit result, P1 to P3 = gaussian amplitude, mean and width, P4 = offset}
  \label{mm_vs_im_improved}
\end{figure}

The result of the cut $|m_X-m_{(K^+,p)}| \le 35$ MeV/c$^2$ is shown in Fig. \ref{mm_vs_im_improved}. The dashed line shows the
invariant mass $m_{(K^+,p)}$ distribution from Fig. \ref{mm_vs_im_3}, while the solid line represents the distribution after we
apply the cut. The peak at $m =$ 1.59 GeV/c$^2$ is our ``$\Theta^{++}$ signal'', which is in good agreement with the prediction of
\cite{BinW03}. The signal-to-background ratio has been improved. The number of events, however, is small. From a fit to the
$\Theta^{++}$ signal with a normal distribution plus a constant offset as background, we get a mass $m_{\Theta^{++}} = m_{(K^+,p)}$
of 1.59 GeV/c$^2$ with a width $\sigma_m$ of 0.01 GeV/c$^2$, $\Gamma_m$ is about 0.025 GeV/c$^2$.

This analysis of the reaction $\gamma$p$\rightarrow$K$^-\Theta^{++}$ preceded the recent experimental discoveries of the $\Theta^+$
by some three years. The $\Theta^+$ mass shows a wide spread between 1.527 GeV/c$^2$ and 1.555 GeV/c$^2$ with a width of up to 0.032
GeV/c$^2$ \cite{Arnd0301}. The current average of the $\Theta^+$ mass from all experiments is about 1.54 GeV/c$^2$. The width
$\Gamma_m$ of 0.025 GeV/c$^2$ for the $\Theta^{++}$ is well within the range seen by other CLAS experiments for the $\Theta^+$
\cite{thet3,thet4,thet7}. The range of 1.5 GeV/c$^2$ to 1.6 GeV/c$^2$ for the $\Theta^++$ search may be too broad, compared to
recent experimental results for the $\Theta^+$ mass, but we also need to take into consideration that the isospin mass splitting for
a pentaquark system with strangeness S=+1 is currently unknown.

\section{The origin of the $\Theta^{++}$ signal}

We now must evaulate the question whether this ``$\Theta^{++}$ signal'' is real or an artifact. It is unlikely that the peak is the
result of an enhancements caused by a combination of the momentum spectra from two particles because of the large acceptance of CLAS
and the fact that we apply only basic cuts. No special cut on the particle momentum is applied. But CLAS does cut off low-momentum
particles. The minimum momentum for particle identification is $p$ {\lower.5ex\hbox{$\stackrel{>}{\sim}$}} 0.2 GeV/c.

Figure \ref{mm_vs_im_1} shows that the signal is far away from most of the mass distributions where the data might be severely
affected by the experimental acceptance and limited total energy. But the mass distributions may be slightly shifted by a few
MeV/c$^2$, due to the momentum-dependent acceptance of CLAS.

Particle misidentification is always an issue. In particular, the separation of pions and kaons with CLAS is difficult. However, the
exclusive measurement and the lack of any missing particle significantly reduce the likelihood of particle misidentification. A pion
misidentified as a kaon would result in a violation of the four-momentum conservation, and there is little phase space left to make
up the difference because of the lack of any undetected heavy particle. If we had, for example, a reaction in which there was a
missing neutron in the final state, then the large missing mass and uncertainty of the momentum measurement would increase the
probability that a particle misidentification remains undetected. Analyses where a kinematical fit is used to constrain the particle
momentum are especially susceptible to such misidentification. This is yet another reason that we chose an exclusive measurement
instead of an inclusive measurement with better statistics.

For this analysis, we do not have to deal with complicated production mechanisms either. There is no spectator and no scattering of
a final-state particle on a spectator or on any other particle. Also Fermi motion is of no concern for this analysis, because the
target is a free proton instead of a deuteron.

But we do have to identify possible reflections from other reactions; there are various reactions possible with a  p, K$^+$, and
K$^-$ in the final state. The obvious candidates are $\gamma$p$\rightarrow$K$^+\Lambda(1520)$,
$\gamma$p$\rightarrow$K$^+\Lambda(1670)$ with $\Lambda^*\rightarrow$K$^-$p and $\gamma$p$\rightarrow\phi$p with
$\phi\rightarrow$K$^+$K$^-$ \cite{PDG02}.

\begin{figure}[ht]
  \epsfbox{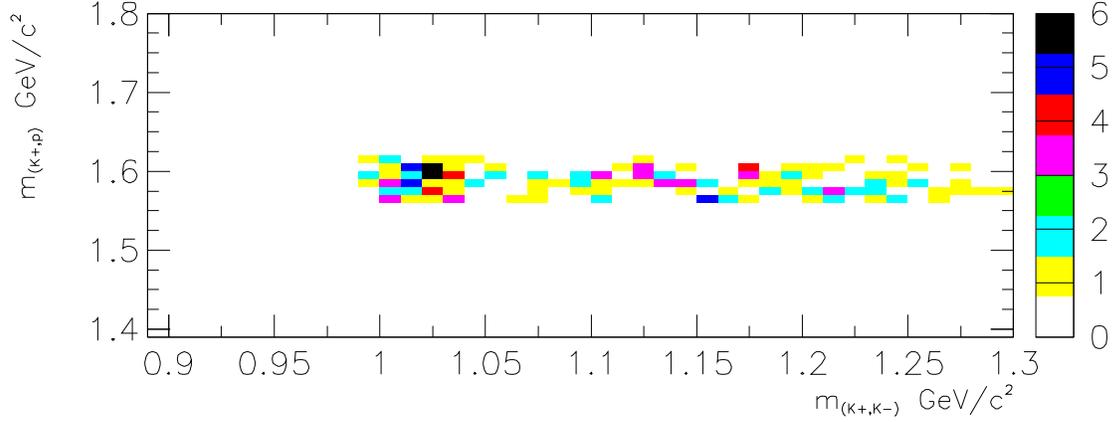}
  \caption{The $\phi$ within the ``$\Theta^{++}$''; invariant mass $m_{(K^+,p)}$ vs. invariant mass $m_{(K^+,K^-)}$, showing the
  correlation between the ``$\Theta^{++}$'' in $m_{(K^+,p)}$ and the $\phi$ in $m_{(K^+,K^-)}$}
  \label{phi_within_zpp_1}
\end{figure}

\begin{figure}[ht]
  \epsfbox{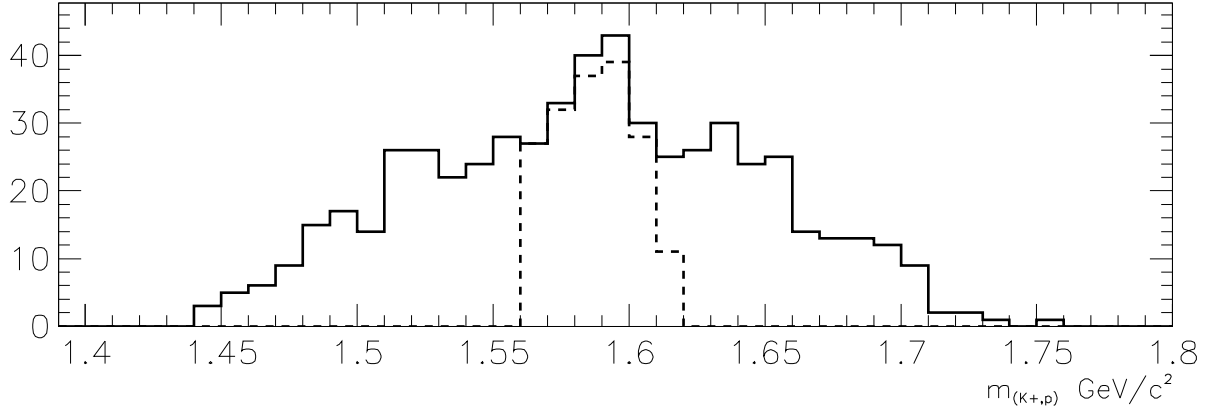}
  \caption{The $\phi$ within the ``$\Theta^{++}$''; same as Fig. \protect\ref{phi_within_zpp_1}, showing the invariant
   mass distribution $m_{(K^+,p)}$ of the selected $\Theta^{++}$ events for Fig. \protect\ref{phi_within_zpp_3} below}
  \label{phi_within_zpp_2}
\end{figure}

\begin{figure}[ht]
  \epsfbox{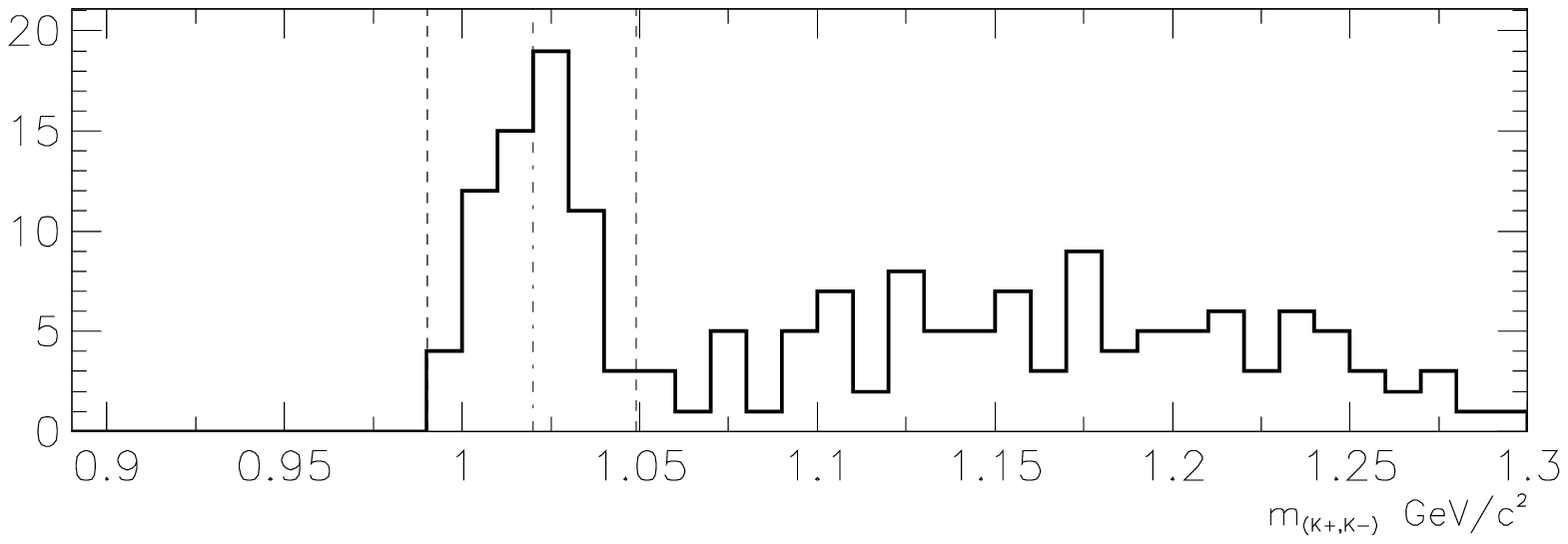}
  \caption{The $\phi$ within the ``$\Theta^{++}$''; same as Fig. \protect\ref{phi_within_zpp_1}, showing the invariant
   mass distribution $m_{(K^+,K^-)}$ for the ``$\Theta^{++}$'' events as indicated in Fig. \protect\ref{phi_within_zpp_2}}
  \label{phi_within_zpp_3}
\end{figure}

First, we search for $\phi$ production in the events within the $\Theta^{++}$ signal. The invariant mass $m_{(K^+,p)}$ versus the
invariant mass $m_{(K^+,K^-)}$ is shown in Fig. \ref{phi_within_zpp_1}. The Y-- and X--projection of Fig. \ref{phi_within_zpp_1} are
shown in Fig. \ref{phi_within_zpp_2} and \ref{phi_within_zpp_3} respectively. We cut on the $\Theta^{++}$ peak, select the
corresponding events below the peak and require $m^2_{(K^+,K^-)}>0$. The selected events are indicated by the dashed line in Fig.
\ref{phi_within_zpp_2}. We then look at the invariant mass $m_{(K^+,K^-)}$, shown in Fig. \ref{phi_within_zpp_3}. Figure
\ref{phi_within_zpp_1}, \ref{phi_within_zpp_2}, and \ref{phi_within_zpp_3} indicate that there is a substantial fraction of $\phi$
production in the group of ``$\Theta^{++}$'' events.

\begin{figure}[ht]
  \epsfbox{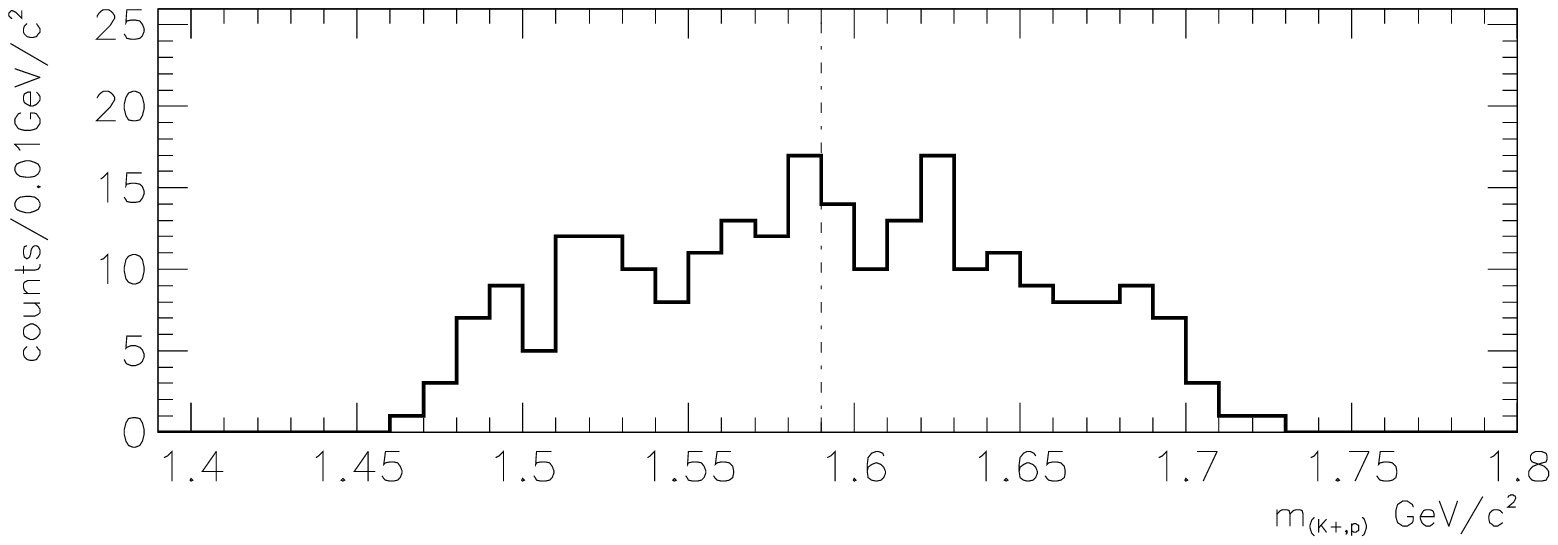}
  \caption{Invariant mass $m_{(K^+,p)}$ for $\gamma$p$\rightarrow$K$^-$K$^+$p, for events with $m_{(K^-,p)} =
   m_{\Lambda(1520)}$}
  \label{zpp_within_y_1}
\end{figure}

\begin{figure}[ht]
  \epsfbox{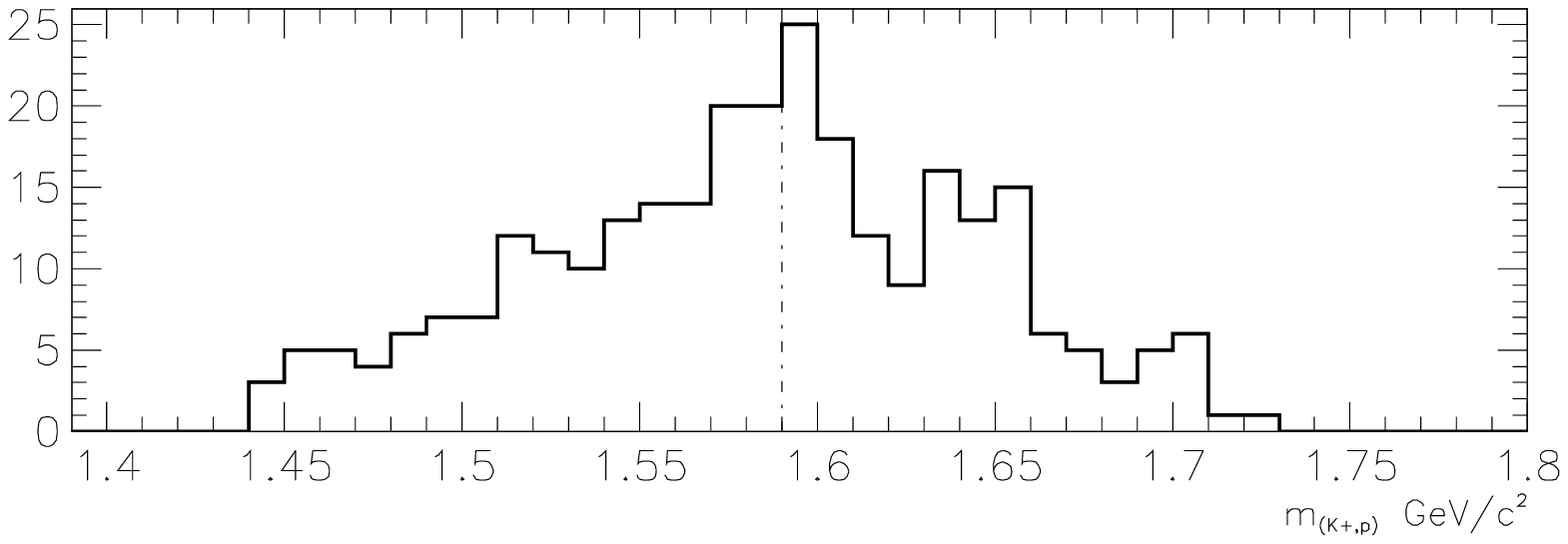}
  \caption{Invariant mass $m_{(K^+,p)}$ for $\gamma$p$\rightarrow$K$^-$K$^+$p, for remainding events after removing the
   hyperon-resonance production, see also Fig. \protect\ref{zpp_within_y_1}}
  \label{zpp_within_y_2}
\end{figure}

To consider hyperon resonance production, the test is also applied the other way around. It turns out that the $\Lambda(1670)$ is of
no importance for this test. Only the results for $\Lambda(1520)$ are shown in Fig. \ref{zpp_within_y_1} and
\ref{zpp_within_y_2}. We apply two cuts, namely $|m_X-m_{(K^-,p)}| < 35$ MeV/c$^2$ for $\gamma$p$\rightarrow$K$^+$X and
$|m_X-m_{(K^+,p)}| < 35$ MeV/c$^2$ for $\gamma$p$\rightarrow$K$^-$X. In Fig. \ref{zpp_within_y_1} and \ref{zpp_within_y_2} we
compare the $\Theta^{++}$ signal in $m_{(K^+,p)}$ for $\Lambda(1520)$ production events and the rest. The ``$\Theta^{++}$'' peak in
Fig. \ref{zpp_within_y_1} from $\Lambda(1520)$ production is very small. Most of the ``$\Theta^{++}$'' signal in
\ref{zpp_within_y_2} comes from events which were not identified as $\Lambda(1520)$ production.

\section{How much of the ``$\Theta^{++}$'' signal is left}

We select events from $\phi$, $\Lambda(1520)$ and $\Lambda(1670)$ production as described before and remove these events from the
``$\Theta^{++}$'' mass distribution. Then we compare the distributions.

\begin{figure}[ht]
  \epsfbox{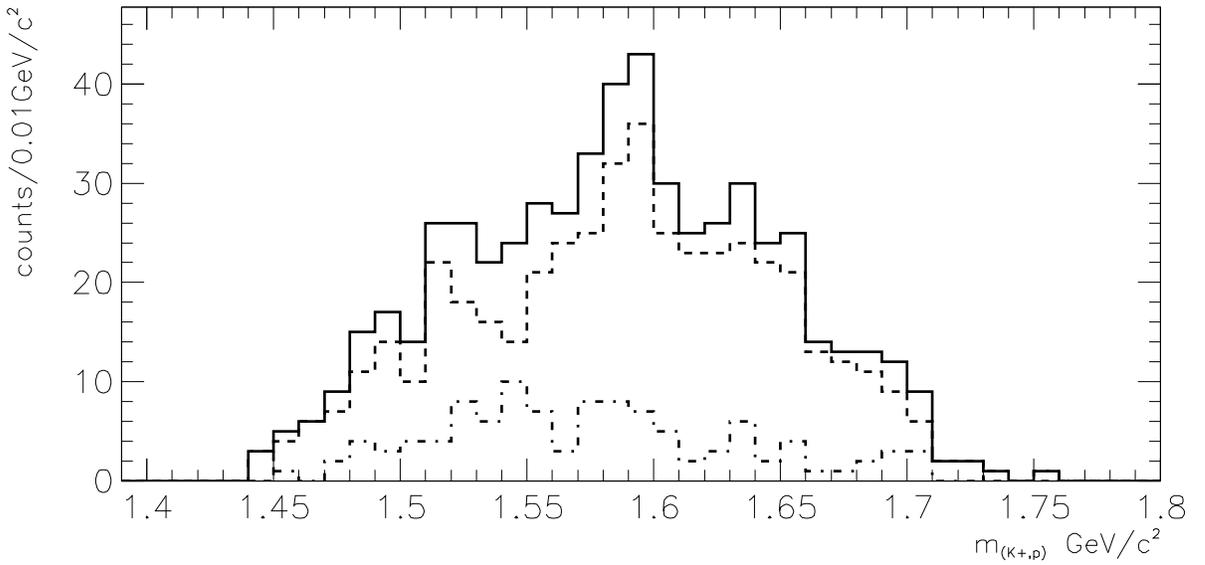}
  \caption{Invariant mass $m_{(K^+,p)}$ of the ``$\Theta^{++}$'' candidates from the original sample as shown in Fig.
   \protect\ref{mm_vs_im_improved} (top line), events coming from either $\phi$, $\Lambda(1520)$ or $\Lambda(1670)$ production
   (middle line) and the difference of both distributions (bottom line)}
  \label{the_rest}
\end{figure}

The top (solid) line in Fig. \ref{the_rest} is again the mass distribution $m_{(K^+,p)}$ as seen in Fig. \ref{mm_vs_im_improved}
with the cut $|m_X-m_{(K^+,p)}| \le 35$ MeV/c$^2$. The middle (dashed) line is the $m_{(K^+,p)}$ distribution from either $\phi$,
$\Lambda(1520)$, or $\Lambda(1670)$ production. The difference between the top and middle distribution plotted as the bottom
(dash-dotted) line of the ``$\Theta^{++}$'' signal, is comparable with a flat line. The signal at 1.59 GeV/c$^2$ is gone. There is
no peak in the range between 1.5 GeV/c$^2$ and 1.6 GeV/c$^2$, in particular not at 1.54 GeV/c$^2$ or 1.59 GeV/c$^2$. The absence of
a $\Theta^{++}$ indicates that the $\Theta^{+}$ state is an isospin-0 isosinglet, in agreement with the conclusions of
\cite{thet5,Aira03}.

\clearpage

\section{Summary}

A peak can be seen in the correlated invariant and missing mass distributions, suggesting a signal from a $\Theta^{++}$ candidate.
However, our analysis shows that this peak can be dismissed as a reflection mainly from $\phi$ production and a smaller contribution
from hyperon-resonance production. Our findings agree with recent results of a $\Theta^{++}$ search in the existing K$^+$-p
scattering data in the W range of 1.520 to 1.560 GeV, which show that there is no evidence for $\Theta^{++}$ with a width greater
than 1 MeV/c$^2$ \cite{Arnd0302}. This suggests that the $\Theta^{+}$ state is an isospin-0 isosinglet.

\end{document}